\documentclass[%
superscriptaddress,
groupedaddress,
reprint,
amsmath,amssymb,
aps,
showpacs,
pra,
]{revtex4-2}

\usepackage{txfonts}
\usepackage{blindtext}
\usepackage{graphicx}
\graphicspath{{figs/}{figsgaoerb/}} 
\usepackage{dcolumn}
\usepackage{bm}
\usepackage{natbib}
\RequirePackage{subcaption}
\usepackage[utf8]{inputenc}
\usepackage[T1]{fontenc}
\usepackage{booktabs, array, mathptmx, float, tabularx, booktabs, lipsum,multirow}
\usepackage{siunitx, xcolor}
\usepackage[colorlinks,linkcolor=blue,anchorcolor=blue,citecolor=blue]{hyperref}

\begin{document}


\title{\textbf{Efficient hybrid variational quantum algorithm for solving graph coloring problem
} 
}%

\author{Dongmei Liu}
 \email{Liudomain@163.com}
\author{Jian Li}%
 \email{lijian@bupt.edu.cn}
\author{Xiubo Cheng}
\affiliation{%
 Key State Laboratory for Network and Switching Technology, Information Security Center, School of Cyberspace Security, Beijing University of Posts and Telecommunications, Beijing, 100876,China
}%

\author{Shibing Zhang}
\author{Yan Chang}
\author{Lili Yan}
\affiliation{
 Cyberspace Security Academy Chengdu University of Information Technology Chengdu, China
}%



\date{\today}

\begin{abstract}\label{abstract}
In the era of Noisy Intermediate Scale Quantum (NISQ) computing, available quantum resources are limited. Many NP-hard problems can be efficiently addressed using hybrid classical and quantum computational methods. This paper proposes a hybrid variational quantum algorithm designed to solve the $k$-coloring problem of graph vertices. The hybrid classical and quantum algorithms primarily partition the graph into multiple subgraphs through hierarchical techniques. The Quantum Approximate Optimization Algorithm (QAOA) is employed to determine the coloring within the subgraphs, while a classical greedy algorithm is utilized to find the coloring of the interaction graph. Fixed coloring is applied to the interaction graph, and feedback is provided to correct any conflicting colorings within the subgraphs. The merging process into the original graph is iteratively optimized to resolve any arising conflicts. We employ a hierarchical framework that integrates feedback correction and conflict resolution to achieve $k$-coloring of arbitrary graph vertices. Through experimental analysis, we demonstrate the effectiveness of the algorithm, highlighting the rapid convergence of conflict evolution and the fact that iterative optimization allows the classical algorithm to approximate the number of colorings. Finally, we apply the proposed algorithm to optimize the scheduling of a subway transportation network, demonstrating a high degree of fairness.

\end{abstract}

\maketitle


\section{Introduction}\label{sec:level1}

The development of quantum computers enables faster computation and greater storage capacity. Many of the challenging classical computational problems today arise from the limited computational speed of current computers, which often results in lengthy processing times to obtain final results. The cost associated with this time consumption is undoubtedly significant, highlighting the urgent need for more efficient solutions. Usually computers are encoded in binary, quantum computer coding depends on the physical medium and can be encoded as N-binary data information \citep{nielsenQuantumComputationQuantum2012}. In the era of noise-containing quantum (NISQ) computing, the number of qubits available to us is still less than ${{10}^{3}}$. Currently, the resources we can use for free are still relatively small, and most of them are teeny qubits of quantum computing resources. Using these computing resources, we can implement and study related quantum computations, such as quantum search algorithms \citep{kockumLectureNotesQuantum2024}, quantum machine learning algorithms \citep{jagerUniversalExpressivenessVariational2023,dingActiveLearningProgrammable2023} quantum adiabatic algorithms \citep{wurtzCounterdiabaticityQuantumApproximate2022,yanAdiabaticQuantumAlgorithm2021,yuExactEquivalenceQuantum2018}, and variational quantum algorithms (VQA) \citep{blekosReviewQuantumApproximate2024,stechlyIntroductionVariationalQuantum2024}, and so on.

At this stage, research on variational quantum algorithms is the most popular. With limited quantum computing resources, we can solve various mappings of NP-hard problems that are difficult to break through by classical computation, such as maximizing \textit{k}-coloring subgraphs, maximal independent subsets, and minimal graph coloring \citep{hadfieldQuantumApproximateOptimization2019}. Variational quantum algorithms are commonly applied to solve the graph max-cut problem, which is one of the easier classes of optimization problems to solve \citep{mollQuantumOptimizationUsing2018, zhouQuantumApproximateOptimization2020, harriganQuantumApproximateOptimization2021}. A general class of quantum lines exists for quantum algorithms for solving the maximal cut problem, where they mix iterations using parametric two-qubit entanglement gates and single-qubit entanglement gates \citep{cerezoVariationalQuantumAlgorithms2021, bhartiNoisyIntermediatescaleQuantum2022}. A series of parameter trainings are then performed in shallow sublines to obtain approximate global optimal parameters \citep{cerezoCostFunctionDependent2021}. Quantum architecture search (QAS) improves the robustness and trainability of VQA, and QAS experiments can be demonstrated to mitigate the effects of quantum noise and barren plateaus \citep{duQuantumCircuitArchitecture2022}. Many scholars tend to study noise elimination and mitigation techniques in quantum circuits in real and simulated environments \citep{wangNoiseinducedBarrenPlateaus2021, endoPracticalQuantumError2018,dingNoiseresistantQuantumState2023,maciejewskiModelingMitigationCrosstalk2021,robbiatiRealtimeErrorMitigation2023,sackLargescaleQuantumApproximate2024}, which can help to increase the number and quality of usable qubits.

In different experimental platforms, we can verify that VQA can outperform classical computation both in terms of quality on solving a particular problem and in terms of computational speed \citep{xuMindSporeQuantumUserfriendly2024,broughtonTensorFlowQuantumSoftware2020}. Quantum computing hardware platforms can be categorized into photonic, ion trap, nuclear magnetic resonance, semiconductor, and superconducting quantum computers \citep{nielsenQuantumComputationQuantum2012}.Google AI's Willow superconducting quantum chip is capable of reaching more than one hundred physical qubits and employs surface error-correcting codes to improve fault tolerance \citep{harriganQuantumApproximateOptimization2021, petersMachineLearningHigh2021,googlequantumaiandcollaboratorsQuantumErrorCorrection2024}.PennyLane offers a unified architecture that optimization and machine learning tasks can be achieved using hybrid classical and quantum implementations \citep{bergholmPennyLaneAutomaticDifferentiation2018}. We use Mindspore Quantum's hybrid quantum classical computing framework \citep{xuMindSporeQuantumUserfriendly2024} to implement a combination of hierarchical quantum approximate optimization algorithms (QAOA) and classical algorithms for solving the \textit{k}-coloring problem on graphs.

The \textit{k}-coloring problem on graphs is a class of NP problems in combinatorial optimization problems. Small-sized graph coloring problems are easy to solve, but for medium to large-sized graphs, the problem size will increase exponentially if classical algorithms are sampled to solve the coloring problem. Common classical algorithms include backtracking \citep{xuResearchAnalysisArtificial2024}, greedy algorithms \cite{sipayungImplementationGreedyAlgorithm2022}, genetic algorithms \cite{ardeleanGraphColoringUsing2022}, neural network algorithms \citep{colantonioEfficientGraphColoring2024}, etc. In 2008, Marco proposed solving the graph coloring problem using a large-scale domain search, which reduces the number of local optimums, but with a large computational cost or time overhead, and does not provide any performance improvement even when incorporating forbidden searches \citep{chiarandiniVeryLargescaleNeighborhood2008}. in 2019, Hu et al. proposed an algorithm to eliminate edge cross-ambiguity in graphs by automatically selecting colors, which is more suitable for small to medium-sized graphs \citep{huColoringAlgorithmDisambiguating2019}.Huang et al. introduced FastColorNet neural network \citep{huangColoringBigGraphs2019}, which aims to solve the coloring problem of large-scale graphs by taking a cue from AlphaGoZero, but the model training relies on a specific dataset, and the generalization ability needs to be improved. In 2022, Schuetz et al. proposed to solve the graph coloring problem with graph neural networks based on the antiferromagnetic Potts model, but the post-processing strategy was simple and could not make full use of the output information of the model \citep{schuetzGraphColoringPhysicsinspired2022}.In 2023, Yannic Maus et al. proposed a deterministic CONGEST algorithm, where the nodes locally compute and try the color sequences, and the improved ruled set algorithm can be applied to distributed graph coloring \citep{mausDistributedGraphColoring2023}.In 2024, Colantonio et al. extended the model to large-scale graphs using GPU parallel computing based on physically-inspired GNN model with better advantages than simulated annealing algorithms \citep{colantonioEfficientGraphColoring2024}.Xu et al. attempted to find a solution by constructing color assignments step by step based on a depth-first search and trial-and-error strategy, but the time complexity is high, especially in dealing with large or high vertex degree graphs with high efficiency \citep{xuResearchAnalysisArtificial2024}.

Classical graph coloring, although algorithmically diverse, exhibits an exponential increase in time complexity for medium to large scale graph coloring. We adopt the QAOA framework of hybrid classical quantum computing, which can achieve polynomial time complexity and explore a wider solution space in a short period of time.In 2019, Oh et al. study VQE and QAOA to solve \textit{k}-coloring on graphs, and research applications to real-world problems, such as boarding gate allocation, frequency allocation, and register allocation problems, but the specific experiments are small in scale \citep{ohSolvingMulticoloringCombinatorial2019}.In 2020, Do et al. focused on the graph coloring problem in quantum algorithm compilation \citep{doPlanningCompilationQuantum2020}, and used time planning for quantum circuit routing, but from the experimental results, the performance of each time planner was not stable. In the same year, Tabi et al. used a spatially efficient graph-coloring embedding method to reduce the number of qubits to $n\left\lceil \log k \right\rceil $ \citep{tabiQuantumOptimizationGraph2020}, but the chosen graph type is simple, and the success rate is relatively low if the coloring is done on complex graphs.

Similarly, there are many NP puzzles using variational quantum algorithms to solve NP puzzles such as coloring on graphs.In 2022, Bravyi they used qudits coding and experiments focused on \textit{d}-regular 3-colored connected graphs to implement QAOA and RQAOA for solving the Max-\textit{k}-cut problem \cite{bravyiHybridQuantumclassicalAlgorithms2022}. Zhou proposed a $QAOA^{2}$ algorithm for solving the large-scale graph MaxCut problem, in which the strategy of partitioning is used \cite{zhouQAOAinQAOASolvingLargescale2022}. In order to improve the quality of the solution, parallel processing of the subgraphs is used by exploiting the ${{\mathbb{Z}}_{2}}$ symmetry. 2023, Deller proposed three methods, namely, penalty term method, conditional gate method, and dynamic decoupling, to solve problems such as graph coloring and electric vehicle charging optimization by QAOA algorithm in qudits system with processing constraints \cite{dellerQuantumApproximateOptimization2023}. In the same year, they similarly focused on solving the Figure 3 coloring problem with \cite{bottrillExploringPotentialQutrits2023} and the QAOA algorithm, and experimentally verified that fewer entanglement gates can be used and the higher the sampling correctness.

The work presented in this paper can be primarily summarized as a systematic approach to hierarchical partitioning, feedback optimization, and conflict resolution. We employ the quantum approximate optimization algorithm (QAOA) in conjunction with classical algorithms to hierarchically address the graph coloring problem. The balanced community detection algorithm is utilized to partition the problem graph into multiple subgraphs. Within these subgraphs, the QAOA algorithm is applied to solve the coloring problem. Additionally, an interaction graph is constructed, and a classical greedy algorithm is employed to determine the coloring of the interaction graph. Subsequently, the results of the interaction coloring are fixed and fed back into the subgraphs to resolve any conflicting colorings. Finally, we address the additional conflicts that arise when merging subgraphs. The paper is organized as follows: In Section \ref{sec:level2},  we discuss color coding, graph coloring conflict loss functions and their associated Hamiltonian. Section \ref{sec:level3} focuses on the implementation principles of the hierarchical hybrid algorithm. In Section \ref{sec:level4}, we present numerical simulation experiments and their application, demonstrating the effectiveness of our algorithm in resolving conflicts. We compare two methods of iterative optimization and non-iterative optimization in terms of coloring success rates and the number of colorings. Additionally, we evaluate our algorithm's performance against other classical algorithms, considering factors such as time, storage, and the number of colorings. Finally, we apply our algorithm to solve real-world problems, including the optimal scheduling of subway transportation networks. Section \ref{sec:level5} concludes the paper.

\section{Preliminary}\label{sec:level2}
\subsection{Color coding}\label{sec:level2.1}
The graph coloring problem is addressed using the quantum algorithm known as QAOA, where the method of color coding significantly influences coloring efficiency. The graph coloring problem (GCP) can be simply described as follows: in a graph $G=(V,E)$, assign a color $c$ to each vertex $v\in V$ such that any edge $(u,v)\in E$ connects vertices of different colors, i.e., $c(u)\ne c(v)$. The objective is to determine the minimum number of colors $k$ required to satisfy this condition. We employ a binary encoding to represent the color of each vertex, $k$ colors stored in the QAOA. The binary string ${{c}_{i}}=({{c}_{i1}},{{c}_{i2}}, ... ,{{c}_{im}})$ encodes colors using a minimum length of $m=\left\lceil {{\log }_{2}}k \right\rceil $ qubits, where ${{c}_{ij}}\in \{0,1\}$.

If the coloring $k=2$, the coloring states (0 and 1) can be represented using a binary number of $m=1$ bit. When the coloring $k=3$, the length is $m=\left\lceil {{\log }_{2}}3 \right\rceil =2$ bits. The binary can represent 4 (00,01,10,11) states, which covers the need for 3 color encodings, leaving one invalid state. For $k=4$, the 2-bit binary representation can exactly accommodate 4 colors, with no resources wasted. In the Mindquantum quantum computing experiment, the maximum number of bits available $Q\approx 21$, and not all of the limited number of qubits can be utilized. After partitioning the subgraphs and initiating the coloring process, we can determine the maximum number of vertices that can be processed in each subgraph.

\begin{equation}
{{V}_{sub}}\le \left\lfloor \frac{Q}{\left\lceil {{\log }_{2}}k \right\rceil } \right\rfloor.\label{eq1}
\end{equation}

When dividing the subgraph, it is essential to limit the number of vertices. Exceeding this constraint may result in an error message indicating that the program is running slowly or that there is insufficient memory allocated. In addition to restricting the number of vertices in a subgraph, edge constraints are necessary to address the \textit{k}-coloring problem for graph vertices. Specifically, two vertices that are adjacent to the same edge must have different color complements.

\subsection{Coloring conflict functions}\label{sec:level2.2}
When graph nodes are assigned colors, it is crucial to minimize the number of colors used while ensuring that adjacent nodes do not share the same color. Within a subgraph, if there are neighboring vertices $u$ and $v$ that have the same color, this situation is classified as a conflict.

\begin{eqnarray}
    conflict(u,v)=\left\{ \begin{matrix}
   1,\exists (u,v)\in E,c(u)=c(v)  \\
   0,otherwise  \\
\end{matrix} \right. \label{eq2}
\end{eqnarray}


We partition the original graph into multiple subgraphs ${{{G}_{1}},{{G}_{2}}, ... ,{{G}_{n}}}$, and construct an interaction graph that represents the interconnections between these subgraphs. In the interaction graph, each subgraph is treated as a supernode, and two supernodes are interconnected whenever an edge exists between their corresponding subgraphs. Naturally, supernodes are prioritized over regular nodes, and conflicts are addressed by fixing the coloring of the interaction graph while adjusting the colors of conflicting vertices within the subgraphs. Conflicts arise when merging multiple subgraphs or when integrating subgraphs with the interaction graph. If there are two subgraphs ${{S}_{i}}$ and ${{S}_{j}}(i\ne j)$, where ${{c}_{{{S}_{i}}}}(u)$ denotes the color of a vertex in subgraph ${{S}_{i}}$ and ${{c}_{{{S}_{j}}}}(u)$ denotes the color of a vertex in subgraph ${{S}_{j}}$, then the associated conflict can be defined as follows:

\begin{eqnarray}
    conflict(u)=\left\{ \begin{matrix}
   1,u\in \bigcup\nolimits_{i\ne j}{(V({{S}_{i}})\bigcap V({{S}_{j}}))},{{c}_{{{S}_{i}}}}(u)\ne {{c}_{{{S}_{j}}}}(u)  \\
   0,\text{otherwise}  \\
\end{matrix} \right.\label{eq3}
\end{eqnarray}

In summary, the total conflict function defined by the trade-off coefficients $\lambda =\left| E \right|/\left| V \right|$, throughout the graph \textit{G}, is:

\begin{eqnarray}
    {{C}_{total}}=\sum\nolimits_{(u,v)\in E}{conflict(u,v)}+\lambda \sum\nolimits_{(u)\in V}{conflict(u)}.\label{eq4}
\end{eqnarray}

If ${{C}_{total}}=0$, it indicates that there is no conflict or the conflict elimination is complete; otherwise ${{C}_{total}}\ne 0$, we need to deal with the conflict further. To conserve computational resources and enhance efficiency, there may be structurally symmetric subgraphs among multiple subgraphs. In such cases, we can apply the same coloring scheme to minimize the total number of colorings or to prevent cross-conflicts. Let ${{\text{G}}_{i}}$ and ${{\text{G}}_{j}}$ represent mutually symmetric sets of graphs,where ${{G}_{i}}\cong {{G}_{j}}$, and let $\phi $ denote the isomorphic mapping $u={{\phi }^{-1}}(v)$ from subgraphs ${{\text{G}}_{i}}\to {{\text{G}}_{j}}$. Then, for all $\forall v\in {{V}_{j}}$, the coloring representation is:

\begin{eqnarray}
    {{c}_{{{G}_{i}}}}(v)={{c}_{{{G}_{j}}}}({{\phi }^{-1}}(v)). \label{eq5}
\end{eqnarray}

In graph coloring problems, an efficient method for mapping and adjusting colors is crucial. This is necessary not only to ensure that neighboring vertices are assigned different colors but also to address the relationship between the interaction graph and the internal coloring of subgraphs. For $n$ subgraphs ${{G}_{1}},{{G}_{2}}, ... ,{{G}_{n}}$, ${{c}_{in}}(v)$ denote the original color of vertex $v$ within the subgraph, and let ${{G}_{i}}$ represent the maximum number of available colors, defined as ${{k}_{i}}={{\max }_{v\in {{G}_{i}}}}{{c}_{in}}(v)+1$. The interaction graph $g'$ consists of the connectivity relations between subgraphs, with the number of colors $k=\chi (g')$ and ${{\alpha }_{i}}={{c}_{base}}({{G}_{i}})$ denotes the base color in the interaction graph $g'$. Color merging encompasses both color remapping and color combining, and the following describes the color synthesis for vertex \textit{v}:

\paragraph{\textbf{Color remapping.}} when ${{k}_{i}}>k$, the number of internal subgraph colors ${{k}_{i}}$ exceeds the number of interaction graph colors $k$, allowing for reuse of colors. However, directly using the original colors may result in color clashes within the subgraph. To achieve color overlay, the internal colors of the subgraph are summed with the base colors, and then taking the mode $k$ of the resulting values can help limit the color space.
\paragraph{\textbf{Color combination.}}When ${{k}_{i}}\le k$, indicating that the number of colors within the subgraph is limited, the base color of the interaction graph can be utilized as the high level, while the internal color of the subgraph serves as the low level. This approach allows for the construction of a composite color and the application of the mode $k\times {{k}_{i}}$ to maintain the boundary. This method helps to avoid conflicts and ensures uniqueness.
\paragraph{\textbf{The new color synthesis rule:}}
\begin{eqnarray}
    {{c}_{\text{mer}ged}}(v)=\left\{ \begin{matrix}
   ({{\alpha }_{i}}+{{c}_{in}}(v))\bmod k,{{k}_{i}}>k  \\
   ({{\alpha }_{i}}\times {{k}_{i}}+{{c}_{in}}(v))\bmod (k\times {{k}_{i}}),{{k}_{i}}\le k  \\
\end{matrix} \right.
\label{eq6}
\end{eqnarray}

Inside the subgraph ${{G}_{i}}$, if the neighboring vertices $(u,v)\in E({{G}_{i}})$ share the same color, the set of conflicting edges $E_{conflict}^{{}}=\{(u,v)\in E|c(u)=c(v)\}$ is modified to eliminate the conflict by changing the color of one of the conflicting vertices, specifically vertex $u$. The objective of color merging is to adjust colors to prevent global edge conflicts, ensuring that the colors within the subgraph remain non-conflicting. The term $M=k\cdot {{\max }_{i}}{{k}_{i}}$ represents the color space modulus, which is utilized to resolve color conflicts between adjacent vertices $u$ and $v$. This is achieved by dynamically adjusting the step size $\Delta (u)=1 + \left\lfloor \frac{c(u)}{M} \right\rfloor $ to address the color conflicts $E_{conflict}^{{}}$ by dynamically adjusting the step size $\Delta (u)=1+\left\lfloor \frac{c(u)}{M} \right\rfloor $:
\begin{eqnarray}
    {{c}_{merged}}(u)=\{(c(u)+\Delta (u))\bmod (M)\}.\label{eq7}
\end{eqnarray}

In the final effect display, we incorporate a validity assessment for coloring and calculate the conflict rate. After merging into the final coloring graph, we verify whether all vertices share the same color to determine the validity of the final coloring scheme. For any vertex $\forall (u,v)\in E$, the validity assessment function of the coloring scheme $c$ is expressed as follows:

\begin{eqnarray}
    Valid(c)=\left\{ \begin{matrix}
1,\text{if}\forall (u,v)\in E,{{c}_{u}}\ne {{c}_{v}}({{c}_{u}}=c_{u}^{(fixed)})  \\
0,otherwise  \\
\end{matrix} \right.
\label{eq8}
\end{eqnarray}

For the entire original graph $G=(V,E)$, assume that the total number of edges is $\left| E \right|$ and the number of conflict edges is $\left| {{E}_{conflict}} \right|=\left| \{(u,v)\in E|c(u)=c(v)\} \right|$. The conflict rate can be computed is expressed as follows:

\begin{eqnarray}
    \varepsilon =\frac{\left| {{E}_{conflict}} \right|}{\left| E \right|}.\label{eq9}
\end{eqnarray}

\subsection{Hamiltonian of the constitutive problem}\label{sec:level2.3}
The QAOA algorithm must be executed multiple times to effectively train for a near-optimal solution. QAOA is applied to address the vertex coloring problem across various subgraphs. However, when the graph is complex or large in scale, utilizing QAOA to solve the coloring of the interaction graph becomes challenging. Consequently, classical methods are employed to color the interaction graph. Once the results of the interaction graph coloring are established, the algorithm proceeds to the interior of the subgraph to determine the corresponding distinct colorings. The constructed problem Hamiltonian primarily ensures that adjacent edges are assigned different colors, thereby establishing the coloring outcome of the interaction graph. This problem Hamiltonian incorporates both soft constraints (neighboring nodes must have different colors) and hard constraints (fixed colors). The QAOA algorithm aims to minimize the expectation of the Hamiltonian, and the resulting optimal quantum state corresponds to the least conflicting coloring of the graph, meaning that neighboring nodes are assigned different colors. For an edge $(u,v)$, the formula for the Hamiltonian quantity related to the soft constraint is expressed as follows:

\begin{equation}
    H_{edge}^{(u,v)}=\sum\limits_{i=0}^{m-1}{(I-{{Z}_{u,i}}\otimes {{Z}_{v,i}})}.\label{eq10}
\end{equation}

${{Z}_{u,i}}$ denotes the Pauli \textit{Z} operator acting on the \textit{i}th qubit of vertex \textit{u}. Since $\text{Z}\left| 0 \right\rangle =\left| 0 \right\rangle ,\text{Z}\left| 1 \right\rangle =-\left| 1 \right\rangle ,$ so the expectation value is given by $\left\langle {{Z}_{u,i}} \right\rangle =1-2{{{b}_{i}}}^{(u)}$. A penalty is incurred if adjacent vertices $(u,v)$ share the same color; otherwise, no penalty is applied. The expectation value indicates that ${{Z}_{u}}{{Z}_{v}}=+1$ in the case of a 00/11 same-color conflict, while it indicates that ${{Z}_{u}}{{Z}_{v}}=-1$ in the case of a 01/10 different-color legality. The conflict penalty factor ${{\lambda }_{edge}}=2$ is designed to amplify the penalty and increase the weight of the conflict term in the optimization process.

The interaction graph is colored to establish the color of the corresponding node, resulting in a fixed-color constrained Hamiltonian quantity $H_{fix}^{(u)}$. Let the color ${{c}^{(u)}}$ be binary encoded as (${{b}_{0}}{{b}_{1}}...{{b}_{m-1}}$), where $b_{i}^{(u)}\in \{0,1\}$ represents the \textit{i}-th position of the binary encoding for the ${{c}^{(u)}}$ target color of vertex \textit{u}. Let $S$ denote the set of fixed vertices in the interaction graph, with penalty term coefficients defined as ${{\lambda }_{fix}}=1000\times \left| S \right|$. The condition ${{\lambda }_{fix}}>{{\lambda }_{\text{ed}ge}}$ ensures that the fixed color constraints take precedence over the edge constraints. For each vertex \textit{u} with a fixed color, the hard constraint Hamiltonian quantity can be expressed as follows:

\begin{eqnarray}
    H_{fix}^{(u)}=\sum\limits_{i=0}^{m-1}{(\left| 2{{b}_{i}}^{(u)}-1 \right|\cdot {{Z}_{u,i}})}+m.
    \label{eq11}
\end{eqnarray}

When the color coding of node \textit{u} is fixed as ${{c}^{(u)}}={{b}_{0}}{{b}_{1}}... {{b}_{m-1}}$(${{b}_{i}}^{(u)}\in \{0,1\}$) when for each Pauli operation ${{Z}_{u,i}}$, $H_{fix}^{(u)}=\sum\limits_{i=0}{(1-2b_{i}^{(u)}){{Z}_{u,i}}}+m$ is a non-strict hard constrained linear approximation representation. When the energy of the system reaches a minimum, this time just corresponds to the qubit encoding of the target color; conversely, if it deviates from the target encoding, the energy value rises. If the color encoding is not fixed, and the color encoding of vertex \textit{u} is $c'\ne {{c}^{(u)}}$ (there exists at least one color binary encoding different from ${{b}_{i}}'\ne {{b}_{i}}$), then $H_{fix}^{(u)}=\sum\limits_{i=0}^{m-1}{(1-2{{b}_{i} })(1-2{{b}_{i}}')}+m\ge 0$. $H_{fix}^{(u)}>0$ means that there are arbitrary bits of color mismatch, and high energy penalty; $H_{fix}^{(u)}=0$ means that the color coding is fixed, and the lower the energy, the closer to the target color. Eventually, the total Hamiltonian quantity ${{H}_{C}}$ and the loss function $L(\theta )$ can be expressed as follows:

\begin{eqnarray}
{{H}_{\text{C}}}={{\lambda }_{edge}}\sum\limits_{(u,v)\in E}{H_{edge}^{(u,v)}+}{{\lambda }_{fix}}\sum\limits_{u\in V}{H_{fix}^{(u)}}.\label{eq12}
\end{eqnarray}

\begin{equation}
	\ell (\theta )=\langle \psi (\theta )|{{H}_{C}}\left| \psi (\theta ) \right\rangle  \\ 
	 =\sum\limits_{(u,v)\in E}{\left\langle H_{edge}^{(u,v)} \right\rangle +}\sum\limits_{u\in G}{\left\langle H_{fix}^{(u)} \right\rangle }.\label{eq13}
\end{equation}

The color \textit{c} is represented as an $m$-bit binary value. The total Hamiltonian employs an energy penalty mechanism: a lower energy value indicates an easier path to obtaining a valid solution, provided there are no conflicts and the color is valid. Conversely, if there is a conflict or an invalid color, the energy value of the conflicting solution is significantly higher. The total Hamiltonian serves as the objective function for optimization, and the selection of key parameters greatly influences the effectiveness of graph coloring. Specifically, it is important to ensure that ${{\lambda }_{fix}}\ge {{\lambda }_{edge}}$ to prioritize fixed colors. By minimizing the loss function $\min (\ell (\theta ))$, we ensure that the measurements of the output states conform to legal coloring, thereby addressing the issues of vertex coloring in the interaction graph and preventing conflicts.

\section{Hierarchical hybrid algorithm}\label{sec:level3}
\begin{figure*}
\centering
\includegraphics[width=0.8\textwidth]{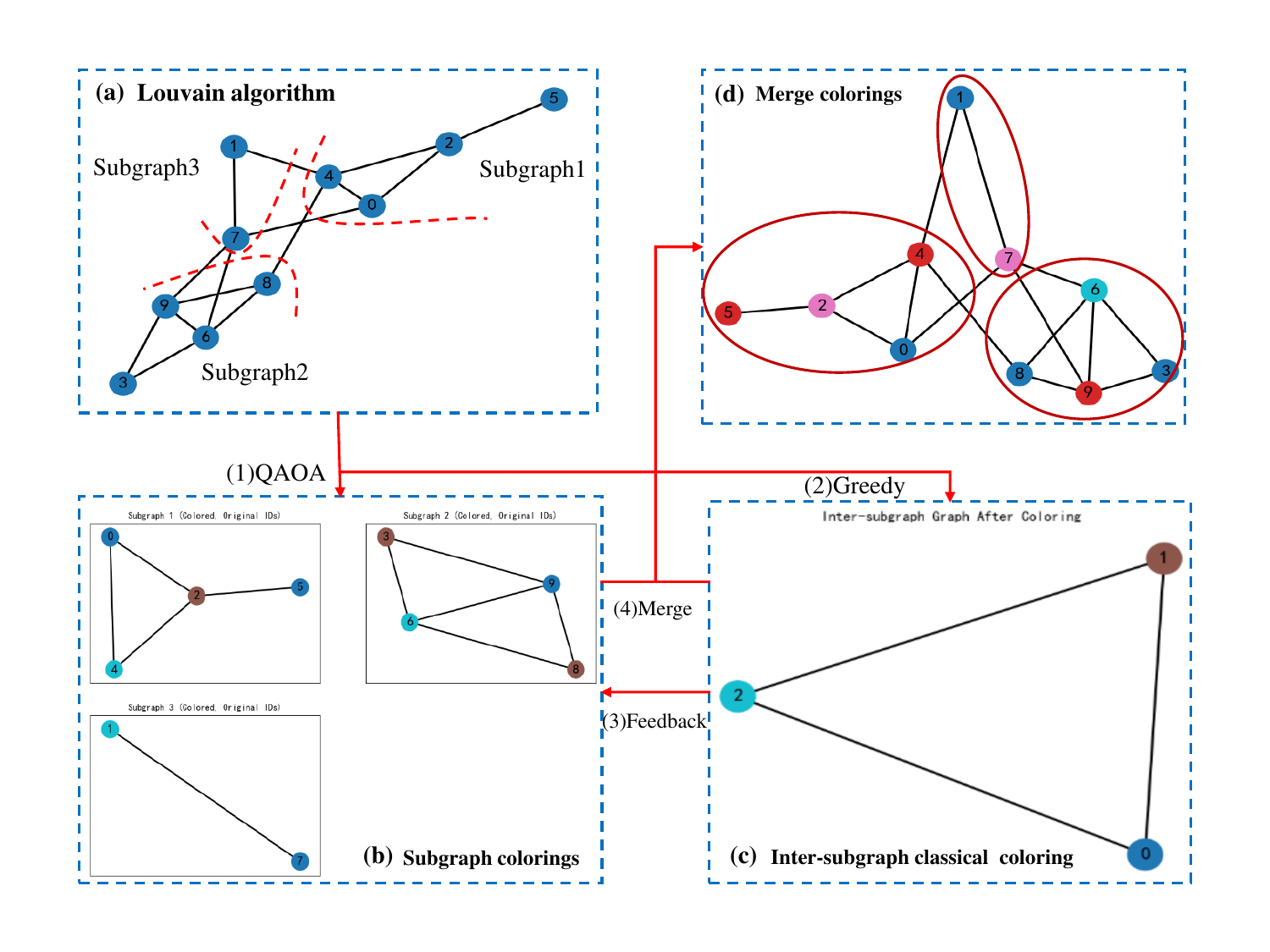}
\caption{\label{fig1}Hierarchical hybrid variational quantum algorithm for solving graph \textit{k}-coloring.}
\end{figure*}

To apply the QAOA algorithm for solveing GCPs in medium to large-scale scenarios, this paper employs hierarchical partitioning and a hybrid variational quantum algorithm. The approach consists of the following steps: (\textit{a}) utilizing the classical Louvain algorithm for balanced community detection to partition the graph into multiple subgraphs, while simultaneously constructing an interaction graph based on the connections between these subgraphs; (\textit{b}) solving the coloring of multiple subgraphs using the QAOA algorithm, with a limitation on the number of qubits used; (\textit{c}) applying a classical greedy algorithm to color the interaction graphs, fixing the coloring of the interaction graphs and feeding the results back to refine the coloring of the subgraphs; (\textit{d}) merging the colorings of the subgraphs and the interaction graph to address any conflicts that arise. The detailed workflow illustrated in Fig.\ref{fig1} below briefly describes the principles behind the 10-node solution for graph coloring.

The work presented in Fig.\ref{fig1} can be mainly summarized as Hierarchical Classification, Feedback Optimization, and Conflict Resolution. This approach employs binary color coding, dynamic penalty constraints, and parameter-adaptive optimization to address the coloring problem. First, there are various types of graphs to choose from, and we select the graph to be colored based on the number of nodes \textit{N} and edges \textit{E}. There are various types of partitioning algorithms available, and Louvain's algorithm for community detection based on modularity is both efficient and scalable. (\textit{a}) The graph employs the Louvain algorithm to partition the original graph, prioritizing nodes with high degree values to be placed in the interior of the subgraph as much as possible during the partitioning process (degree centrality-based partitioning). This approach minimizes conflicts encountered during the merging phase. Additionally, a dynamic method for determining the number of partitioned subgraphs, defined as ${{n}_{s}}=\max (2,\sqrt{N})$ is utilized to ensure that each subgraph can be colored as effectively as possible. Subsequently, the number of nodes in each subgraph is arranged in descending order, with subgraph 1 containing the largest number of nodes, which aids in conserving computational resources.

Process (1) refers to the QAOA algorithm applied to each subgraph to determine the optimal result for subgraph coloring. The specific workflow for the (\textit{b}) subgraph colorings depicted in Fig.\ref{fig1} is shown in Fig.\ref{fig2} below. In Fig.\ref{fig2}, it is essential to check whether the subgraph is empty or to dynamically limit the size of the subgraph nodes within a reasonable range of QAOA computational resources. Subsequently, each subgraph is traversed sequentially by the QAOA algorithm to identify the near-optimal solution for \textit{k}-coloring. If a subgraph requires more qubits than the available computational resources to solve the coloring problem, it is skipped. Otherwise, the final output consists of the number of colorings and the coloring corresponding to the smallest \textit{k}-value.

\begin{figure*}
\centering
\includegraphics[width=0.8\textwidth]{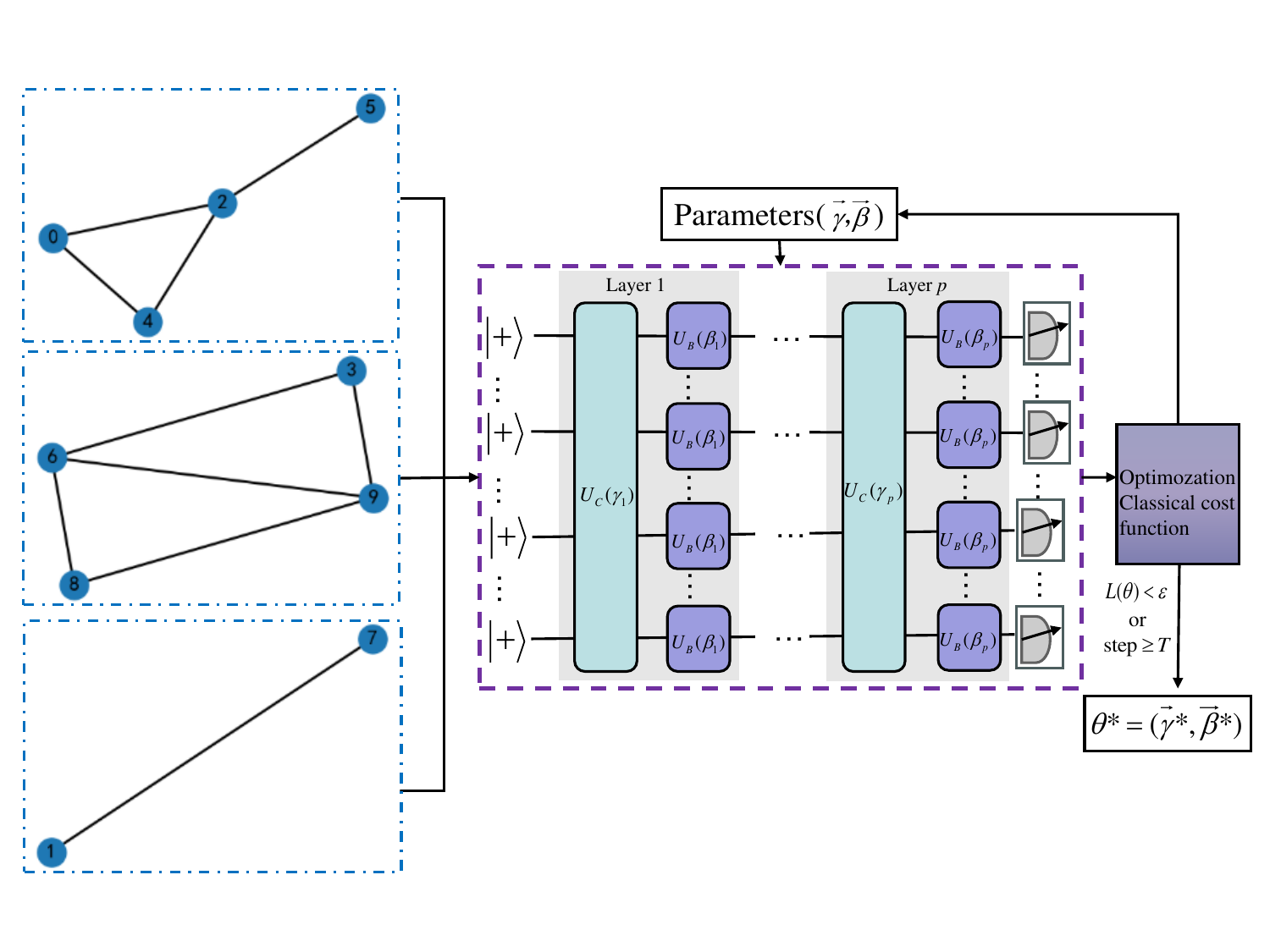}
\caption{\label{fig2}Subgraphs in QAOA are used to solve graph coloring \textit{k} problems.}
\end{figure*}

Fig.\ref{fig2} refers to the workflow for solving the subgraph \textit{k}-coloring problem using QAOA. This process includes constructing the ansatz circuit, initializing the simulator, formulating the Hamiltonian, optimizing the parameters(optimal angles), generating samples, and extracting the corresponding sample coloring results. Parameter optimization necessitates quantum optimization and simulator emulation, with the algorithm designed to produce a subgraph coloring scheme that minimizes conflicts by incrementally testing different \textit{k} values. Since the sampling results from the QAOA algorithm are based on an approximation of the minimum Hamiltonian rather than an exact zero-conflict solution, some subgraphs may be colored with a few conflicts, while others may yield an optimal solution. The pseudo-code for the \textit{k}-coloring algorithm is provided in detail in Table\ref{tab:table1} below.

A cache is utilized throughout the hierarchical process to store the results of subgraph coloring. During the post-processing stage, multiple conflicting edge colorings may arise when merging subgraphs back into the original graph. In accordance with the graph vertex \textit{k}-coloring requirement that no two adjacent vertices can share the same color, we must address these conflicting edges. As the size of the graph increases, the number of divided subgraphs also rises, potentially causing the conflict domain to expand either linearly or exponentially. If the QAOA algorithm is employed to resolve the conflicting edge colorings of the subgraphs, it will conserve more resources compared to processing global coloring conflicts. However, since the QAOA has already been applied to complete the subgraph coloring, we will enhance the convergence rate by implementing a straightforward conflict-handling method, specifically local greedy conflict-correction coloring, which can ultimately lead to convergence or the resolution of conflicts.

In Fig.\ref{fig1}, we employ the classical greedy algorithm outlined in step (2) to address the interactive graph vertex coloring problem. However, if the number of edges in the interaction graph is excessively large, the classical greedy algorithm may struggle with complex coloring challenges. Step (3) is implemented to refine the interaction graph coloring, optimize the feedback effects, and resolve any previous conflicting QAOA colorings within the subgraph. Finally, step (4) merges the subgraphs and resolves the coloring conflicts between them to achieve the original graph coloring, as illustrated in Fig.\ref{fig1} (\textit{d}).

The optimization iteration function in process (3) mainly implements feedback corrective coloring of the graph. This process includes initializing the cache, processing (\textit{b}) subgraph coloring (specifically symmetric graph coloring), merging colors, detecting conflicts, and adjusting colors in iterations to resolve these conflicts. While we have previously described the subgraph coloring process in detail, we will now refer to section \ref{sec:level2.2} and focus on the color merging and conflict handling part. Fig.\ref{fig1} (\textit{c}) after the coloring of the interaction graph is completed, the coloring results of the subgraph are continuously optimized through a color space expansion strategy and a conflict feedback adjustment mechanism until a globally valid coloring is achieved or the maximum number of iterations is reached. Subsequently, the subgraph coloring results, referred to as sub\_colorings, along with the interaction graph coloring scheme, known as inter\_coloring, are passed as parameters to the merge coloring function in process (4). The merge coloring function is primarily responsible for merging the optimized subgraph coloring results to resolve conflicts and create the final graph coloring scheme. During the merging process, it dynamically adjusts the parameters of each subgraph based on a caching mechanism to avoid redundant calculations and address conflicts between subgraphs. Additionally, it allows for monitoring the evolution of the number of conflicting edges, as illustrated in Fig.\ref{fig3} below. Ultimately, the original graph coloring results are produced.

Processes (3) and (4) are post-processing operations illustrated in Fig.\ref{fig1}, which focus on managing color conflicts. In the iterative optimization process (3), color conflicts within the subgraphs are addressed using a conflict feedback adjustment mechanism. Meanwhile, process (4) deals with inter-subgraph connection conflicts and the global conflicts that arise when merging the coloring results of the subgraphs. Although the levels of conflict addressed differ, both processes ensure the validity of the graph coloring scheme. Each process requires adjustments to the subgraph color assignments through strategies for color space expansion and color compression mapping. Ultimately, our algorithm assigns the minimum number of colors to the graph vertices and effectively eliminates color conflicts.

\begin{table}[b]
	\caption{\label{tab:table1}%
		Solving subgraph \textit{k}-coloring pseudo-code.}
	\begin{ruledtabular}
		\begin{tabular}{lcdr}
			QAOA algorithm \textit{solve\_k\_coloring}:\\
			Input: graph \textit{G}, maximum number of colors ${{K}_{\max }}$, number of QAOA \\layers \textit{p}\\
			Output: minimum number of colors $k*$ and legal coloring \textit{C}\\
			\colrule
			\textit{a}. Quantum bit allocation: ${{n}_{qubits}}=\max (2,\left\lceil {{\log }_{2}}k \right\rceil )$,\\${{N}_{total}}=\left| V \right|\cdot {{n}_{qubits}}$, if ${{N}_{total}>Q}$, then jump out.\\
			\textit{b}. Construct the Hamiltonian quantity ${{H}_{C}}$:\\ ${{H}_{\text{C}}}={{\lambda }_{edge}}\sum\limits_{(u,v)\in E}{H_{edge}^{(u,v)}+}{{\lambda }_{fix}}\sum\limits_{u\in V}{H_{fix}^{(u)}}$.\\
			\textit{c}. QAOA ansatz construction: initialize quantum state\\ $\left| {{\psi }_{0}} \right\rangle ={{\left| + \right\rangle }^{\otimes {{N}_{qubits}}}}$, ${{H}_{M}}=\sum\limits_{v\in V}{\sum\limits_{i=0}^{{{n}_{qubits}}-1}{{{X}_{v,i}}}}$, \textit{p}-layer parameter line\\ evolution: $\left| \psi (\overrightarrow{\gamma },\overrightarrow{\beta }) \right\rangle =\prod\limits_{l=1}^{p}{({{e}^{-i{{\beta }_{l}}{{H}_{M}}}}{{e}^{-i{{\gamma }_{l}}{{H}_{C}}}})}\left| {{\psi }_{0}} \right\rangle $.\\
			\textit{d}. Parameter optimization: $\ell (\overrightarrow{\gamma },\overrightarrow{\beta })=\left\langle  \psi (\overrightarrow{\gamma },\overrightarrow{\beta }) \right|{{H}_{C}}\left| \psi (\overrightarrow{\gamma },\overrightarrow{\beta }) \right\rangle $, \\Gradient Updates: ${{\theta }_{t+1}}={{\theta }_{t}}-\eta {{\nabla }_{\theta }}\ell (\theta )$. If the loss value does not\\ improve in 3 consecutive steps or the number of conflict edges\\ trained is 0, stop early.\\
			\textit{e}. With measurement coloring extraction: when the optimal\\ parameter $\theta *=(\overrightarrow{\gamma }*,\overrightarrow{\beta }*)$, measure the quantum final state\\ ${s=({{s}_{0}},{{s}_{1}},...,{{s}_{{{N}_{qubits}}-1}})\sim {{\left| \left\langle  s | \psi (\theta *) \right\rangle  \right|}^{2}}}$ , node \textit{v} color \\extraction: ${{c}_{v}}=(\sum\limits_{i=0}^{{{n}_{qubits}}-1}{{{2}^{i}}{{s}_{v,i}}})\bmod k$.\\
			\textit{f}. Verify legitimacy: $\text{Valid}\Leftrightarrow \left\{ \begin{matrix}
				{{c}_{u}}\ne {{c}_{v}}\forall (u,v)\in E  \\
				{{c}_{u}}=c_{u}^{(fixed)}\forall u\in (fixednodes)  \\
			\end{matrix} \right.$\\
			\colrule
			Output: if a legal solution is found, Return($k*,C={{\{{{{c}_{v}}}\}}_{v\in V}}$)

		\end{tabular}
	\end{ruledtabular}
\end{table}

\section{Experiments and Application}\label{sec:level4}
\subsection{Evolution on the edge of conflict}\label{sec:level4.1}
In QAOA coloring, we employ the hierarchical QAOA method to achieve graph vertex coloring, and the post-processing phase necessitates the elimination of conflicting edges. We have analytically determined that the evolution of conflicting edges changes dynamically; however, the overall conflict can be resolved in just a few steps during the optimization iteration process. During the initial coloring phase, we assign the same initial color to each vertex, meaning that all vertices are initially colored identically. Consequently, the number of edges in the graph at this stage corresponds to the number of conflicting edges; in other words, the original graph contains the maximum number of conflicting edges.

After completing the hierarchical subgraph QAOA coloring and interaction graph coloring, we proceed to the merging and conflict resolution phase. The interaction graph may exhibit cross-subgraph vertex coloring conflicts among the subgraphs. Fixed supernode colors are reintroduced into the subgraph to rectify the identical coloring of the corresponding edges. This corrected coloring may lead to the emergence of new conflicts, which can subsequently be reduced, or the conflicts may diminish rapidly until they reach convergence. The nature of these varying conflict representations is influenced by the complexity of the graph structure. In the case of sparse graphs, the QAOA algorithm operates more efficiently, exploring different \textit{k}-value colorings and managing conflicts, often resolving all conflicts within a few steps. Fig.\ref{fig3} below illustrates the evolution of conflict elimination across different instances.

\begin{figure*}[!t] 
	\centering
	\begin{subfigure}{0.4\textwidth}
		\centering
		\includegraphics[height=2in]{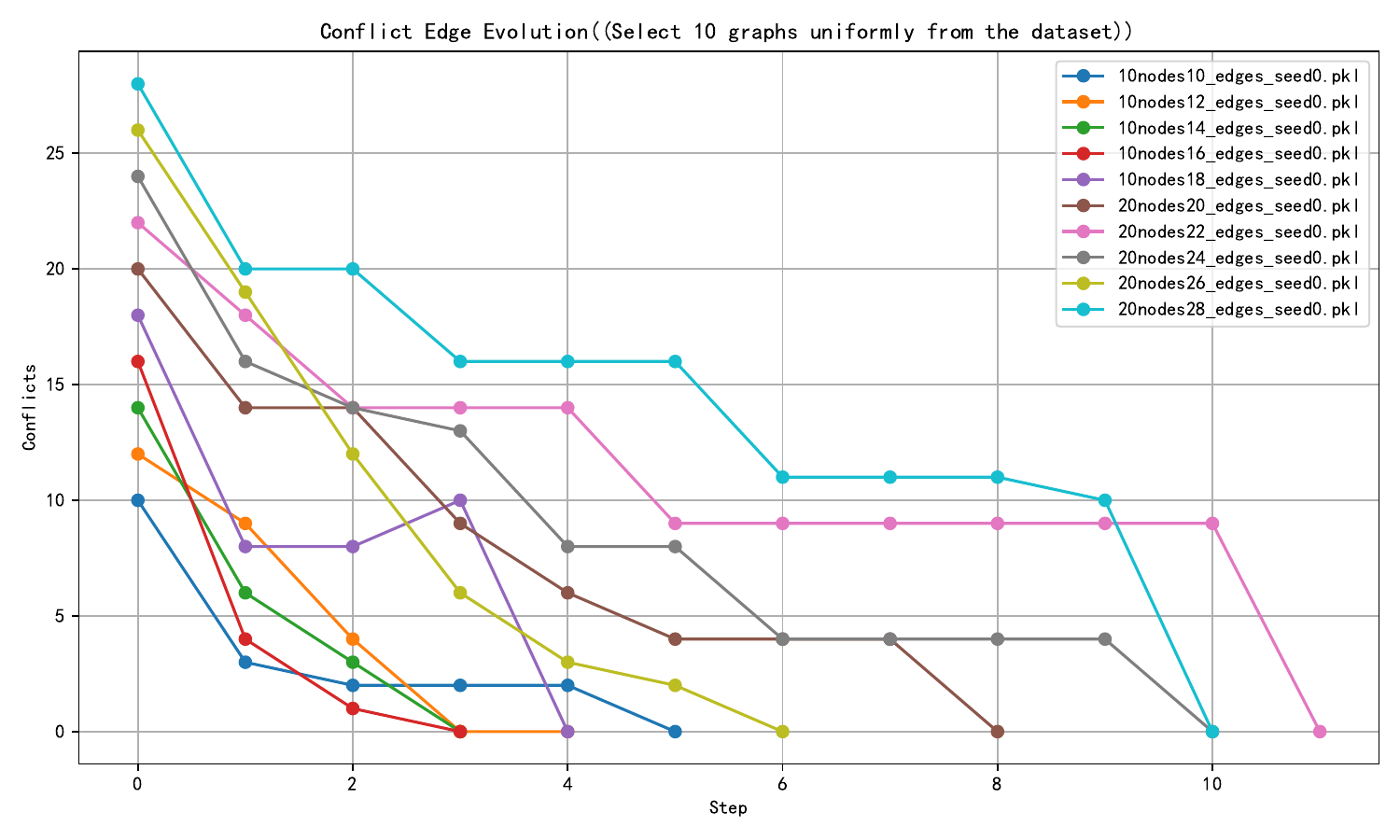}
		\caption{Conflict elimination evolution of 10 graph examples.}
		\label{fig3.1}
	\end{subfigure}
	\hfill
	\begin{subfigure}{0.5\textwidth}
		\centering
		\includegraphics[height=2.2in,width=3.5in]{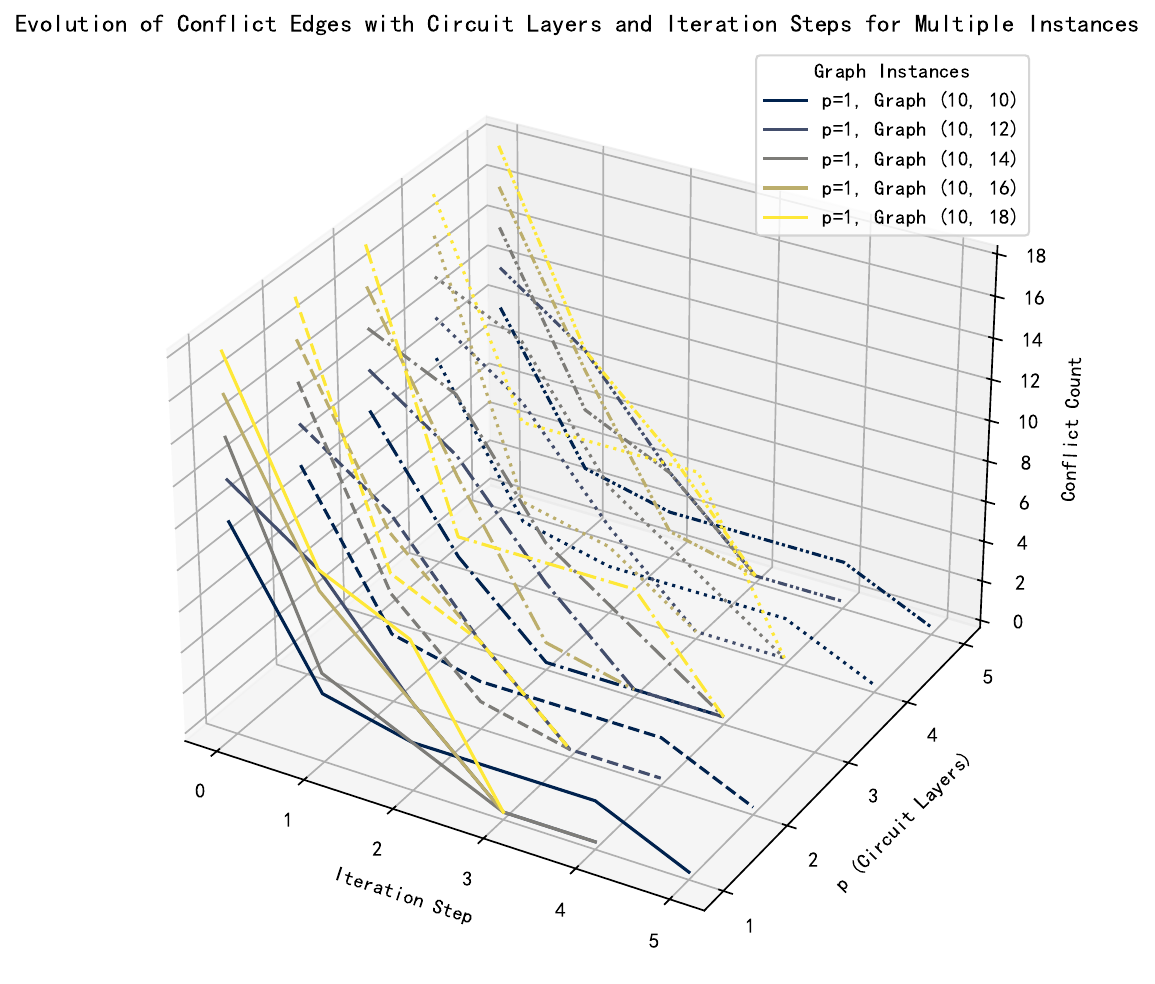}
		\caption{5 examples of different layers of conflict evolution.}
		\label{fig3.2}
	\end{subfigure}
	\caption{Iterative evolution of conflict edges}
	\label{fig3}
\end{figure*}

When the graph structure is simpler, our conflict elimination process evolves more rapidly; conversely, when the graph structure is more complex, the number of steps required for conflict elimination increases. Due to the limitations of platform qubits, the number of qubits constrains both the number of graph coloring vertices and the number of colorings, ultimately affecting overall solving efficiency. For instance, with 20 qubits and \textit{k} = 8, the number of qubit encodings that can be represented per node is $\left\lceil {{\log }_{2}}8 \right\rceil =3$, which limits the number of subgraph nodes that can be processed to $\left\lfloor 20/3 \right\rfloor =6$. When the size of the problem exceeds what can be simulated, we may experience slow simulation speeds, uncertain parameter optimization, or a tendency to converge on local optimal solutions. Therefore, when the graph structure is sparse, it is advisable to begin training with simple parameter values, such as \textit{k} = 2. This approach not only conserves quantum computing resources but also facilitates the rapid identification of the minimum legal coloring.

The algorithm still has some limitations. Since it addresses k-coloring, it must evaluate each k value sequentially, starting from 2. Consequently, when the selected k-values fall within a narrow range, the algorithm may continue to operate with a limited number of conflicts. This situation can lead to inefficient use of training resources, resulting in waste. Additionally, the presence of conflicts creates obstacles for the algorithm during the optimization and iteration processes, significantly slowing down convergence. This delay prolongs the time required to achieve a converged state, ultimately impacting the overall efficiency and performance of the algorithm. With advancements in computational resources, and in accordance with Brooks' theorem, the arbitrary coloring number will not exceed the maximum degree, denoted as $\Delta +1$. We can enhance the algorithm by implementing a continuously decreasing hot-start scheme, such as setting $k = \Delta +1$, in conjunction with a robust strategy for identifying the minimum k-value for coloring. This approach will consider both accuracy and efficiency. 

\subsection{Iterative optimization}\label{sec:level4.2}
The primary objective of the iterative optimization strategy for graph coloring is to minimize the consumption of quantum resources by employing a subgraph division approach, thereby constructing a globally effective legal graph coloring scheme. Step (3) in Fig.\ref{fig1} illustrates a multi-round feedback iterative optimization method, which is utilized to rectify conflicting color assignments within subgraphs by stabilizing the interaction graph coloring. The central concept is to harmonize color conflicts between subgraphs through the use of intermediate graphs. The iterative feedback optimization method encompasses several key features, in addition to multiple rounds of iteration and conflict resolution. These features include utilizing a cache to store intermediate results, sharing a coloring scheme between symmetric subgraphs, employing a classical greedy algorithm to address the interaction graph coloring, merging subgraph color numbering, and verifying the legitimacy of the final result. Through these mechanisms, iterative feedback optimization ensures the legitimacy and accuracy of the solution.

Non-iterative feedback optimization is more convenient and faster than the iterative feedback optimization method. The primary concept involves directly and linearly merging subgraphs after subgraph coloring, without utilizing a feedback mechanism for coordination. Key features of this approach include strategies such as bypassing inter-subgraph coordination and conflict resolution, coloring each subgraph only once, and caching global linear coloring. While this method significantly enhances computational efficiency, it does come at the cost of reduced accuracy in the results.

\begin{figure*}
	\centering
	\includegraphics[width=1.0\textwidth]{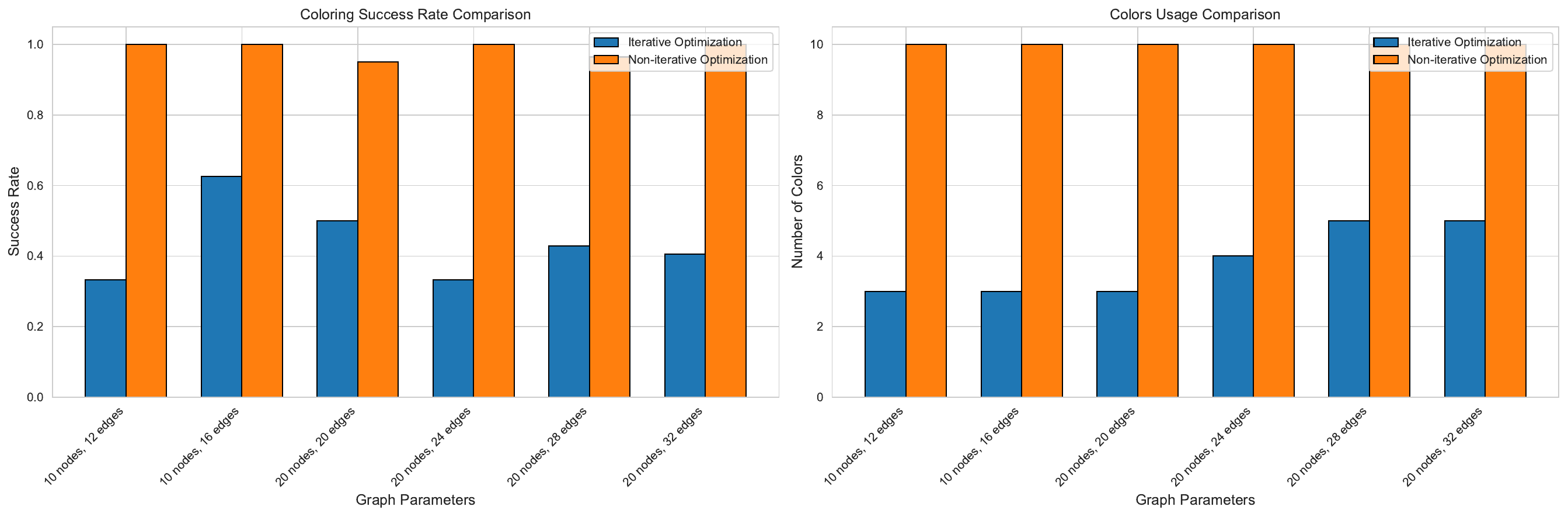}
	\caption{\label{fig4}Performance of iterative and non-iterative optimization methods under different graph structures.}
\end{figure*}

In the left panel of Fig.\ref{fig4}, we compare the success rates of subgraph coloring for iterative and non-iterative optimization. The non-iterative optimization demonstrates a remarkably high success rate on small to medium-density graphs, indicating strong robustness. In contrast, the success rate of iterative optimization is generally below 50\%, particularly underperforming on large graphs or those with dense edges. The final performance metrics reveal a success rate of 43.77\% for iterative optimization and 98.57\% for non-iterative optimization.

Although the success rate of the iterative optimization coloring method is not high, the right panel of Fig.\ref{fig4} compares the number of colors used by both methods for the same graph structure. While the iterative optimization method has a low success rate, it is significantly more advantageous in terms of the number of colors utilized, as it consistently requires fewer colorings than the non-iterative optimization method. As the size of the graph increases, the number of colors needed for iterative optimization also rises. We believe that the underlying reason for this phenomenon is that non-iterative optimization methods typically employ a maximum upper bound on the number of colors (without compression), sacrificing spatial efficiency for the sake of coloring success. In contrast, iterative optimization methods nearly always achieve an approximate optimal number of colors when coloring is successful, making them suitable for resource-sensitive quantum coding application. Ultimately, the average number of colors used by the iterative optimization method is 3.83, while the average number of colors used by the non-iterative optimization method remains consistently at 10.00.

\subsection{Algorithm Performances}\label{sec:level4.3}
\begin{figure*}
	\centering
	\includegraphics[width=1.0\textwidth]{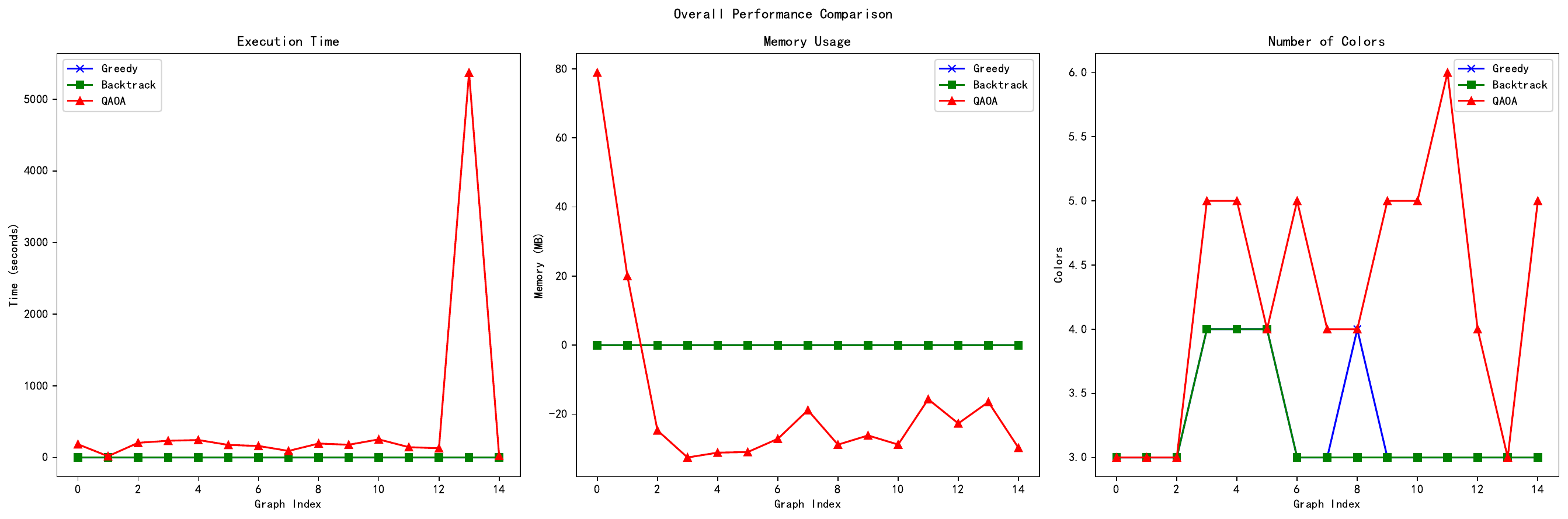}
	\caption{\label{fig5}Comparative analysis of algorithm performance in small-scale graph scenarios.}
\end{figure*}

We compare the hierarchical QAOA algorithm and the classical algorithms (e.g., backtracking, greedy) in solving the graph vertex coloring problem. Our analysis focuses primarily on running time, storage overhead, and the number of colors used in small graph scenarios, as illustrated in Fig.\ref{fig5}. In the two subgraphs depicting execution time and memory usage, it is evident that both the running time and memory consumption of the greedy algorithm and the backtracking algorithm are steadily converging towards zero. These two classical algorithms operate on a simple graph structure, allowing for the rapid identification of optimal solutions. The greedy algorithm makes a locally optimal choice based on the current state, requiring no backtracking, resulting in minimal computation and low time overhead. Conversely, the backtracking algorithm explores the solution space using depth-first search; it backtracks when it cannot find a satisfactory solution and releases memory once the search is complete. In cases of simple graph structures, the backtracking algorithm can quickly eliminate invalid paths to identify the optimal solution. However, when dealing with complex graph types, extensive backtracking may be necessary, leading to a significant increase in running time.

The QAOA algorithm explores multiple \textit{k}-values to identify the minimum coloring. Its execution time tends to be comparable to that of classical algorithms when the problem size is small and the graph structure is simple. However, when addressing large-scale, complex-structured graphs for coloring, the QAOA algorithm requires more time to execute due to the increased complexity of quantum state preparation and evolution operations. Our hierarchical QAOA algorithm utilizes memory space to retain the results of intermediate computations while performing graph vertex coloring. The memory metric shown in Fig.\ref{fig5} is a relative measure, and the initial drop in memory indicates the active release of memory after the completion of the QAOA execution. Subsequently, the memory usage associated with solving \textit{k}-coloring for various graph instances fluctuates with the problem size.

The final subgraph of Fig.\ref{fig5} shows the minimum number of colorings identified through a comparison of the three algorithms. The QAOA algorithm explores the optimal solution by leveraging quantum properties, and the number of colors it employs is influenced by the evolution of the quantum state and the constraints of the problem. Variations in graph structures result in differing quantum state evolutions and probabilistic sampling outcomes, which in turn lead to slight discrepancies in the number of colors required to achieve the optimal coloring scheme.

\subsection{Algorithmic Application}\label{sec:level4.4}
In this paper, we implement various types of \textit{k}-coloring problems based on a collaborative framework that integrates hierarchical partitioning, feedback optimization, and conflict resolution. Building on this foundation, we innovatively adapt the algorithmic model to the scheduling context of the Beijing subway transportation network, establishing a mapping relationship of “graph theory \textit{k}-coloring problem - subway station conflict coloring”. Specifically, the coloring conflict problem of neighboring nodes in graph theory is accurately aligned with the spatio-temporal conflicts of trains operating within the same station or track resources. This alignment aims to prevent safety hazards arising from the simultaneous operation of multiple trains within the same spatio-temporal domain. By implementing the optimized station coloring scheme, we can enhance resource scheduling and operational management of subway lines, significantly improving transportation efficiency. Additionally, by integrating train running times, passenger flow data, and fault emergency response data, we can further mitigate traffic congestion, thereby achieving intelligent management of the subway system.

\begin{figure*}[!t] 
	\centering
	\begin{subfigure}{0.6\textwidth}
		\centering
		\includegraphics[height=3.5in]{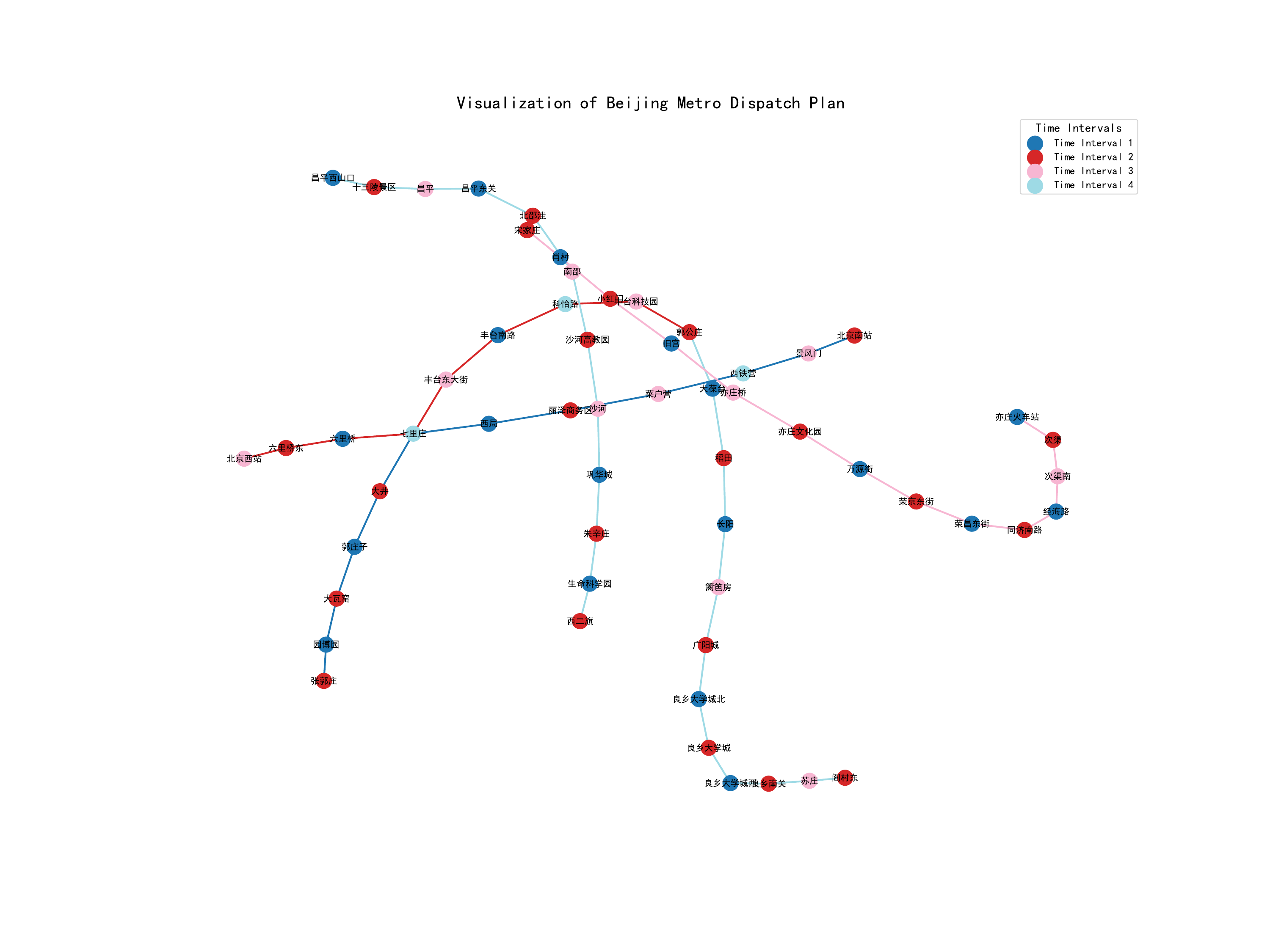}
		\caption{Subway traffic \textit{k} coloring scheduling program.}
		\label{fig6_a}
	\end{subfigure}
	\hfill
	\begin{subfigure}{0.3\textwidth}
		\centering
		\includegraphics[height=2in,width=1.5in]{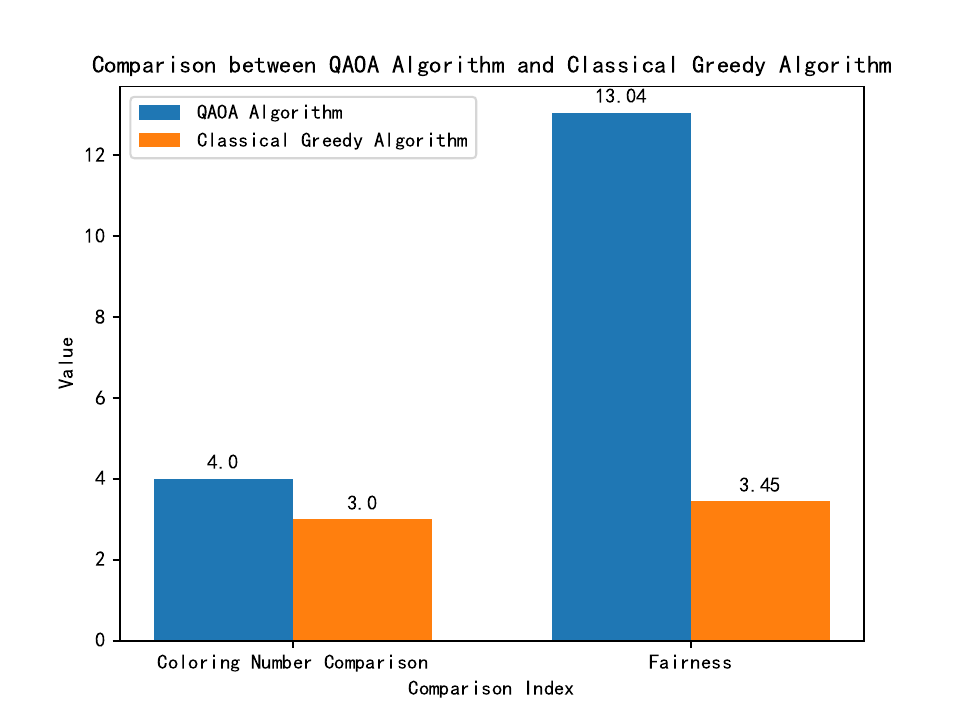}
		\caption{Comparative analysis of metro transit coloring numbers and fairness.}
		\label{fig6_b}
	\end{subfigure}
	\caption{Metro transportation network scheduling and analysis}
	\label{fig6}
\end{figure*}

Fig.\ref{fig6_a} specifically selected segment of the Beijing subway transportation system and its \textit{k}-coloring scheme. The colors on the subway traffic map correspond to specific scheduling time intervals, with the goal of minimizing the number of colors used in the \textit{k}-coloring. Each time interval for a given station falls within the same batch or time period, and the color assigned is determined based on the line to which it belongs, making it easy to distinguish between different lines. Fig.\ref{fig6_b} compares the QAOA algorithm coloring and the classical greedy algorithm of the minimum number of colors used. The two algorithms show little difference in the number of colors required, as illustrated in Fig.\ref{fig5}(c), where their effects appear similar. Fairness is defined as the ratio of the minimum to the maximum number of nodes corresponding to each color in the overall color set. If the maximum value is greater than 0, the fairness is calculated and returned; otherwise, it returns 0. In the fairness index shown in Fig.\ref{fig6_b}, the QAOA algorithm achieves a fairness value of 13.04\%, while the classical greedy algorithm has a fairness value of 3.45\%. The QAOA algorithm's balanced resource allocation significantly outperforms that of the classical greedy algorithm, effectively minimizing resource waste.

Of course, the metro network in the current scenario encompasses a vast amount of data, with numerous nodes and edges. The station layout and line planning are continuously updated to adapt to changing conditions. To effectively manage operating costs while ensuring that train operations can flexibly respond to real-time fluctuations in passenger demand, it is essential to develop a more sophisticated and dynamic scheduling plan through comprehensive research and multi-dimensional optimization.

\section{Conclusion}\label{sec:level5}
This paper investigates the application of variational quantum algorithms to address the graph vertex coloring problem. The primary approach involves seamlessly integrating a classical algorithm with a quantum algorithm, utilizing hierarchical division to partition the graph into multiple subgraphs. Within these subgraphs, the QAOA algorithm is employed to tackle the coloring problem, while the classical greedy algorithm is applied to the interaction graph to determine the coloring scheme. Subsequently, the coloring results from the interaction graph are fixed and incorporated back into the subgraphs to refine their coloring. During the merging of subgraphs, any conflicts that arise are resolved through iterative optimization.

The entire workflow can be summarized in three key steps: hierarchical division, feedback optimization, and conflict resolution. Additionally, color compression and remapping, dynamic penalty constraints, and parameter adaptive optimization are employed to effectively address the subgraph coloring problem. Experimental analysis confirms that the hybrid variational quantum algorithm for solving graph vertex coloring can rapidly eliminate conflicts in edge correction coloring. The iterative evolution within the algorithm minimizes the number of graph colorings, even though the coloring success rate in the subgraph may not be high. In terms of running time, memory usage, and the number of colorings, the QAOA algorithm shows comparable performance to classical algorithms regarding the minimum number of colorings, despite some limitations. Finally, the present algorithm is also applied to the metro transportation network, where successful coloring of the transportation network map can optimize the scheduling scheme.

\begin{acknowledgments}
This work was supported by the National natural Science Foundation of China (Grant No. 62271070), the Open Research Fund of the  Key Laboratory of Cryptography of Zhejiang Province (Grant No. ZCL21006), and the Graduate Innovation Fund of Beijing University of Posts and Telecommunications (Grant No. CX20241061).
\end{acknowledgments}

\bibliographystyle{apsrev4-2}  
\bibliography{reference}  

\end{document}